\documentstyle[emulateapj,apjfonts,psfig,natbib]{article}
\def\swift{{\it Swift}}
\def\nod{\nodata}

\def\ociw{1}
\def\prince{2}
\def\hubble{3}
\def\cit{4}
\def\srl{5}
\def\tapir{6}
\def\uh{5}
\def\anu{6}
\def\nrao{7}
\def\lco{8}
\def\pom{9}
\def\toronto{10}
\def\tok{11}
\def\uq{12}

\begin{document}

\title{\large The Afterglows, Redshifts, and Properties of \swift\ 
Gamma-Ray Bursts}

\author{
E.~Berger\altaffilmark{\ociw,}\altaffilmark{\prince,}\altaffilmark{\hubble},
S.~R.~Kulkarni\altaffilmark{\cit},
D.~B.~Fox\altaffilmark{\cit},
A.~M.~Soderberg\altaffilmark{\cit},
F.~A.~Harrison\altaffilmark{\srl},
E.~Nakar\altaffilmark{\tapir},
D.~D.~Kelson\altaffilmark{\ociw},
M.~D.~Gladders\altaffilmark{\ociw},
J.~S.~Mulchaey\altaffilmark{\ociw},
A.~Oemler\altaffilmark{\ociw},
A.~Dressler\altaffilmark{\ociw},
S.~B.~Cenko\altaffilmark{\cit},
P.~A.~Price\altaffilmark{\uh},
B.~P.~Schmidt\altaffilmark{\anu},
D.~A.~Frail\altaffilmark{\nrao},
N.~Morrell\altaffilmark{\lco},
S.~Gonzalez\altaffilmark{\lco},
W.~Krzeminski\altaffilmark{\lco},
R.~Sari\altaffilmark{\tapir},
A.~Gal-Yam\altaffilmark{\cit,}\altaffilmark{\hubble},
D.-S.~Moon\altaffilmark{\srl},
B.~E.~Penprase\altaffilmark{\pom},
R.~Jayawardhana\altaffilmark{\toronto},
A.~Scholz\altaffilmark{\toronto},
J.~Rich\altaffilmark{\anu},
B.~A.~Peterson\altaffilmark{\anu},
G.~Anderson\altaffilmark{\anu},
R.~McNaught\altaffilmark{\anu},
T.~Minezaki\altaffilmark{\tok},
Y.~Yoshii\altaffilmark{\tok},
L.~L.~Cowie\altaffilmark{\uh},
and K.~Pimbblet\altaffilmark{\uq} 
}

\altaffiltext{\ociw}{Observatories of the Carnegie Institution
of Washington, 813 Santa Barbara Street, Pasadena, CA 91101}
 
\altaffiltext{\prince}{Princeton University Observatory,
Peyton Hall, Ivy Lane, Princeton, NJ 08544}
 
\altaffiltext{\hubble}{Hubble Fellow}

\altaffiltext{\cit}{Division of Physics, Mathematics and Astronomy,
105-24, California Institute of Technology, Pasadena, CA 91125}

\altaffiltext{\srl}{Space Radiation Laboratory, MS 220-47, California 
Institute of Technology, Pasadena, CA 91125}

\altaffiltext{\tapir}{California Institute of Technology, Theoretical
Astrophysics and Relativity Group, MC 130-33, Pasadena, CA 91125}

\altaffiltext{\uh}{Institute for Astronomy, University of Hawaii, 
2680 Woodlawn Drive, Honolulu, HI 96822}

\altaffiltext{\anu}{Research School of Astronomy and Astrophysics, 
Australian National University, Mt Stromlo Observatory, via Cotter Rd,
Weston Creek, ACT 2611, Australia}

\altaffiltext{\nrao}{National Radio Astronomy Observatory, Socorro,
NM 87801}

\altaffiltext{\lco}{Las Campanas Observatory, Carnegie Observatories, 
Casilla 601, La Serena, Chile}

\altaffiltext{\pom}{Pomona College Department of Physics and Astronomy, 
610 N. College Avenue, Claremont, CA}

\altaffiltext{\toronto}{Department of Astronomy and Astrophysics, 
University of Toronto, 60 St. George Street, Toronto, ON M5S 3H8, 
Canada}

\altaffiltext{\tok}{Institute of Astronomy, School of Science, University 
of Tokyo, 2-21-1 Osawa, Mitaka, Tokyo 181-0015, Japan}

\altaffiltext{\uq}{Department of Physics, University of Queensland, 
Brisbane, 4072 Queensland, Australia}

\begin{abstract} 
We present optical, near-IR, and radio follow up of sixteen \swift\ bursts,
including our discovery of nine afterglows and a redshift determination for
three.  These observations, supplemented by data from the literature,
provide an afterglow recovery rate of $60\%$ in the optical/near-IR, much
higher than in previous missions (BeppoSAX, HETE-2, INTEGRAL, and IPN).  
The optical/near-IR afterglows of \swift\ events are on average $1.7$ mag
fainter at $t=12$ hr than those of previous missions.  The X-ray afterglows
are similarly fainter compared to those of pre-\swift\ bursts.  In the radio
the limiting factor is the VLA threshold and the detection rate for \swift\
bursts is similar to that for past missions.  The redshift distribution of
pre-\swift\ bursts peaked at $z\sim 1$, whereas the five \swift\ bursts with
measured redshifts are distributed evenly between 1.3 and 3.2.  From these
results we conclude that (i) the pre-\swift\ distributions were biased in
favor of bright events and low redshift events, (ii) the higher sensitivity
and accurate positions of \swift\ result in a better representation of the
true burst redshift and brightness distributions (which are higher and
dimmer, respectively), and (iii) as many as $1/3$ of the bursts can be
optically dark, as a result of a high redshift and/or dust extinction.  We
remark that the apparent lack of low redshift, low luminosity \swift\
bursts, and the lower event rate compared to pre-launch estimates (90 
vs.~150 per year), are the result of a threshold that is similar to that 
of BATSE.  In view of these inferences, afterglow observers may find it 
advisable to make significant changes in follow up strategies of 
\swift\ events.  The faintness of the afterglows means that large 
telescopes should be employed as soon as the burst is localized.  
Sensitive observations in $RIz$ and near-IR bands will be needed to 
discriminate between a typical $z\sim 2$ burst with modest extinction and 
a high redshift event.  Radio observations will be profitable for a small 
fraction ($\sim 10\%$) of events.  Finally, we suggest that a search for 
bright host galaxies in untriggered BAT localizations may increase the 
chance of finding nearby low luminosity GRBs.
\end{abstract}

\keywords{gamma-rays:bursts}

\section{Introduction}
\label{sec:intro}

The \swift\ satellite \citep{gcg+04} has now been operational for over
four months.  So far, about two dozen bursts have been rapidly
localized to better than $10''$ accuracy thanks to the on-board X-ray
telescope (XRT).  Such precise and rapid positions are critical for
deep ground-based follow up, in particular for the determination of
redshifts and whether a burst is optically ``dark'' (due to extinction
within the host galaxy or attenuation by Lyman scattering from
intergalactic hydrogen).

For past missions (BeppoSAX, HETE-2/WXM, INTEGRAL, IPN) while $\sim
90\%$ of the afterglows were detected in the X-rays
\citep{pir01,bkf03,dpp+03}, the fraction with optical and
radio afterglow (essential for arcsecond localization) was only $30\%$
(e.g., \citealt{fjg+01,lcg02,bkb+02,fkb+03}).  Bursts localized by the
Soft X-ray Camera (SXC; \citealt{vvd+99}) on-board HETE-2, on the
other hand, had a recovery rate in the optical of about $85\%$ and in
the radio of about $55\%$ \citep{lra+04,bfk+05}.  This has been
attributed to the relatively accurate and rapid positions provided by
the SXC, and has placed the tightest limit on the fraction of dark
bursts.

The current \swift\ sample of well localized bursts is now of
sufficient size to provide a meaningful comparison to past missions,
and to start to draw statistical inferences about the GRB population.
Of particular interest is whether the increased sensitivity of \swift\
results in a sizable fraction of low redshift, low luminosity GRBs, or
an increase in the detection rate of GRBs at higher redshifts.  A
related question is whether the recovery rate of optical/near-IR
afterglows is close to unity.

Here we present our comprehensive optical/near-IR and radio follow-up
observations of \swift\ bursts.  Of the 21 \swift\ GRBs with XRT
positions and ground-based follow up we observed a total of sixteen in
the optical/near-IR and thirteen in the radio.  We discovered nine of
the twelve optical/near-IR afterglows to date, radio afterglows for
three bursts, and determined redshifts of three \swift\ bursts and one
SXC burst.  We show that the optical/near-IR detection rate for
\swift\ bursts is indeed higher than in past missions, but that the
afterglows are significantly fainter, and their redshifts tend to be
higher.  Deep limits suggest that as many as $1/3$ of the \swift\
bursts can be optically dark.  These conclusions have major
ramifications for future follow up efforts, which we discuss towards
the end of the paper.

\section{Afterglow Discovery and Redshifts}
\label{sec:obs}

Follow-up observations by our group were made using an armada of
telescopes at the following facilities: Las Campanas Observatory
(LCO), Magellan, Palomar Observatory, Keck, Siding Spring Observatory
(SSO), and the Very Large Array (VLA\footnotemark\footnotetext{The VLA
is operated by the National Radio Astronomy Observatory, a facility of
the National Science Foundation operated under cooperative agreement
by Associated Universities, Inc.}).  Afterglow discovery and follow up
of GRBs 041223, 050117a, 050124, and 050126 were detailed in
\citet{bfk+05}.  In Table~\ref{tab:obs} we provide photometry and
radio flux measurements for subsequent events, augmented by relevant
data from the literature.  

All optical/near-IR observations were reduced in the standard manner
using IRAF routines.  Astrometry was performed relative to the USNO-B
catalog, and the afterglow identifications were made by comparison to
the Digitized Sky Survey (DSS) or through detection of a fading
behavior.  The VLA radio observations were undertaken in the standard
continuum mode, and the data were reduced and analyzed using the
Astronomical Image Processing System.

\begin{figure}
\centerline{\psfig{file=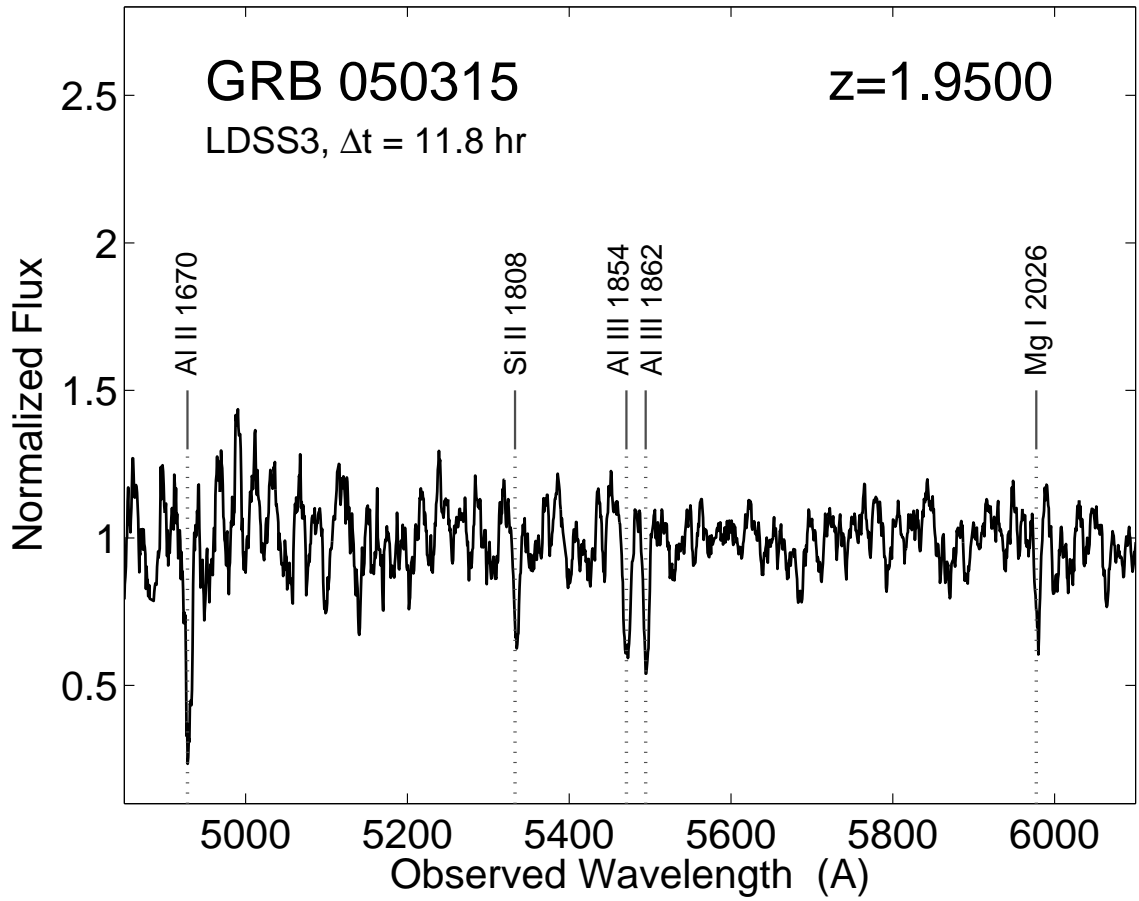,width=6.0in}}
\caption{Absorption spectrum of GRB\,050315 taken with the LDSS3
instrument on the Magellan/Clay 6.5-m telescope 11.8 hr after the
burst ($R=20.9$ mag; Table~ref{tab:obs}).  The spectrum exhibits
several absorption features corresponding to \ion{Al}{2}\,($\lambda
1670$), \ion{Si}{2}\,($\lambda 1808$), \ion{Al}{3}\,($\lambda \lambda
1854,1862$), and \ion{Mg}{1}\,($\lambda 2026$) at a redshift,
$z=1.9500\pm 0.0008$.
\label{fig:g050315}}
\end{figure}

\begin{figure}
\centerline{\psfig{file=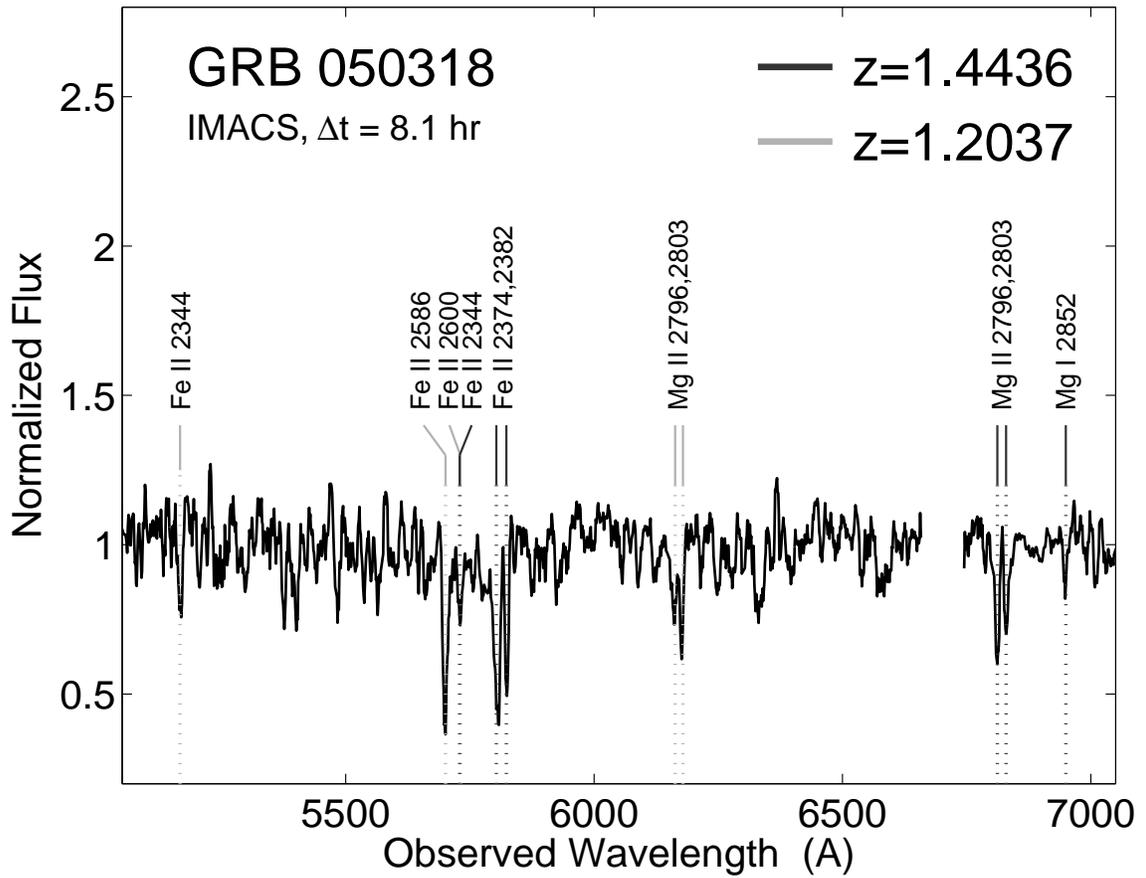,width=6.0in}}
\caption{Absorption spectrum of GRB\,050318 taken with the IMACS
instrument on the Magellan/Baade 6.5-m telescope 8.1 hr after the
burst ($R=20.6$ mag; Table~ref{tab:obs}).  The spectrum exhibits
several absorption features corresponding to Fe and Mg lines at
redshifts, $z_1=1.2037\pm 0.0004$ and $z_2=1.4436\pm 0.0009$.  We
identify the higher redshift system with GRB\,050318.
\label{fig:g050318}}
\end{figure}

\begin{figure}
\centerline{\psfig{file=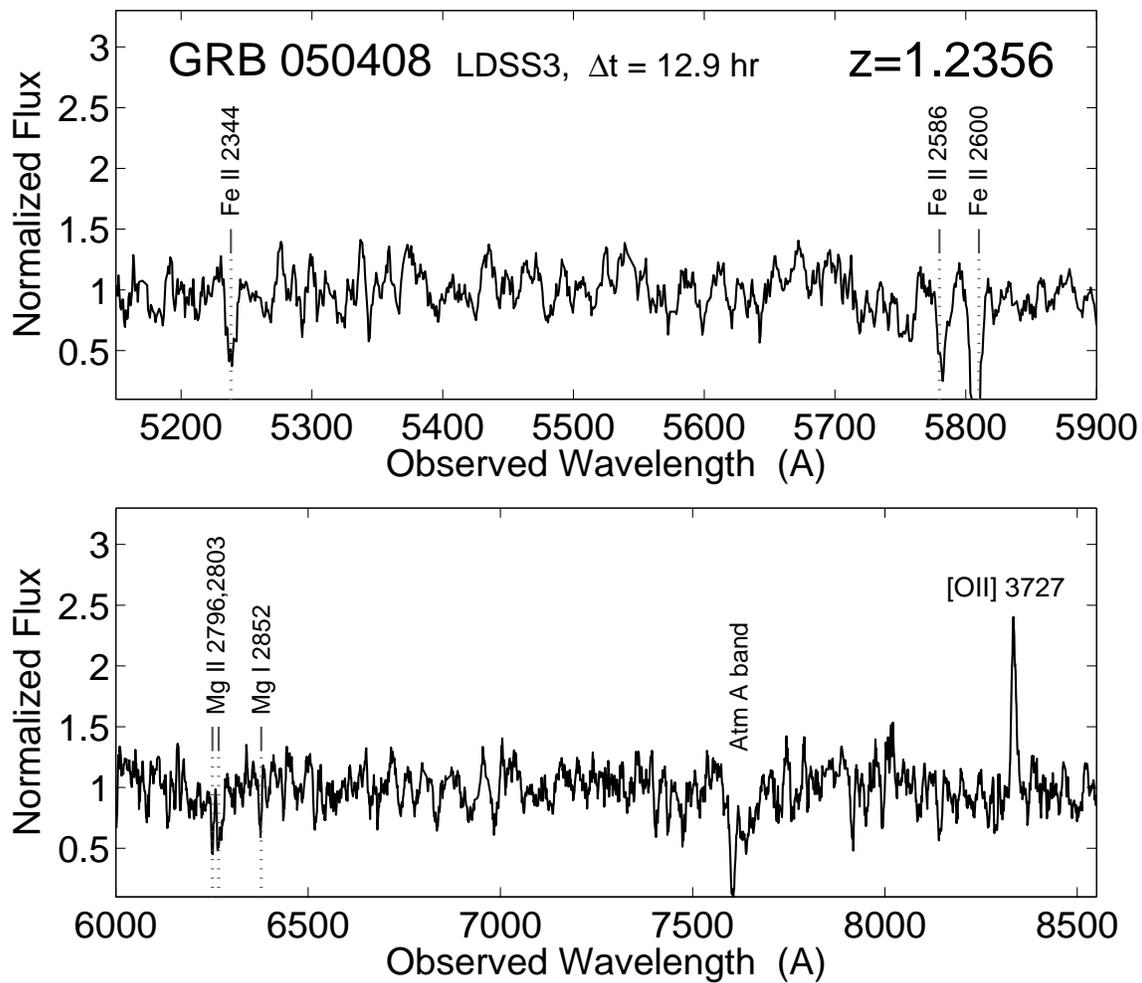,width=6.0in}}
\caption{Absorption spectrum of the HETE-2 SXC burst GRB\,050408
taken with the LDSS3 instrument on the Magllan/Clay 6.5-m telescope
about 12.9 hr after the burst ($R=22.0$ mag).  The spectrum exhibits
absorption from Fe and Mg, as well as an [OII]$\lambda 3727$ emission
line.  The redshift of the burst is $z=1.2356\pm 0.0008$.
\label{fig:g050408}}
\end{figure}

We used the LDSS3 and IMACS spectrographs on the Magellan 6.5-m
telescopes to obtain absorption spectra for the \swift\ GRBs 050315
(Figure~\ref{fig:g050315}) and 050318 (Figure~\ref{fig:g050318}), and
the HETE-2 SXC burst GRB\,050408 (Figure~\ref{fig:g050408}).  We also
used ESI on the Keck-II telescope to obtain a redshift for the host
galaxy of GRB\,050126 \citep{bfk+05}.  In all cases, we used standard
IRAF routines to bias-subtract and flat-field the data, while
rectification and sky subtraction were performed using the method and
software described in \citet{kel03}.  Air-to-vacuum and heliocentric
corrections were applied to the wavelength calibration.  The redshifts
determined from these spectra are listed in Table~\ref{tab:obs}, along
with two other redshifts available from the literature.  A detailed
analysis of the absorption spectra will be provided in a separate
paper (Berger {\it et al.} in prep.)

Finally, we list in Table~\ref{tab:obs} the X-ray fluxes and
$\gamma$-ray peak photon fluxes and fluences when available from the
literature.  For GRBs 050401, 050406, 050416a, and 050421 we undertook
our own analysis of the XRT data (from the \swift\
archive\footnotemark\footnotetext{\sf
http://heasarc.gsfc.nasa.gov/docs/swift/archive/}).  We cleaned the
data using the standard settings in {\sf xrtpipeline}, and extracted
photons in the $0.5-7$ keV band.  This optimizes detection
signal-to-noise for the average afterglow, which has a photon spectral
index of about $-2$.  For the photon counting mode we used an
extraction radius of 20 pixels ($90\%$ encircled energy) for the
source, and an annulus outside of this, starting at a radius of 30
pixels, for background extraction.  The data were fit with a power law
plus absorption.  Finally, we used the measured photon spectral index
to extrapolate the flux to the $2-10$ keV band for comparison with
bursts from other missions.  The conversion from count rate to flux is
$1\,{\rm cps}=2\times 10^{-11}$ erg cm$^{-2}$ s$^{-1}$ ($2-10$ keV).

\section{The Properties of \swift\ Bursts}
\label{sec:prop}

In this section we summarize the properties of the sample of 21
\swift\ bursts that have XRT positions and ground-based
optical/near-IR follow up (Table~\ref{tab:obs}; as of the end of April
2005).  We compare the \swift\ sample with two previous samples: (i)
HETE-2 SXC bursts with positional accuracy better than $2'$ (``SXC''),
and (ii) bursts localized by other past missions (BeppoSAX, HETE-2
WXM, IPN, and INTEGRAL; ``BWI'').  The former sample (14 objects)
enjoys superior localizations, while the latter sample (96 objects)
has moderate localizations ($\sim 3-30'$).

The overall detection fraction of X-ray afterglows for \swift\ bursts
is $21/22$ (one burst detected in $\gamma$-rays has no XRT detection).
This is essentially the same as the detection fraction for past
missions of about $90\%$.  The recovery fraction of optical/near-IR
afterglows for the \swift\ sample, $12/21\approx 60\%$, is
significantly higher than the $30\%$ recovery fraction of the BWI
sample, but is slightly worse than the $85\%$ fraction for the SXC
bursts.

\begin{figure}
\centerline{\psfig{file=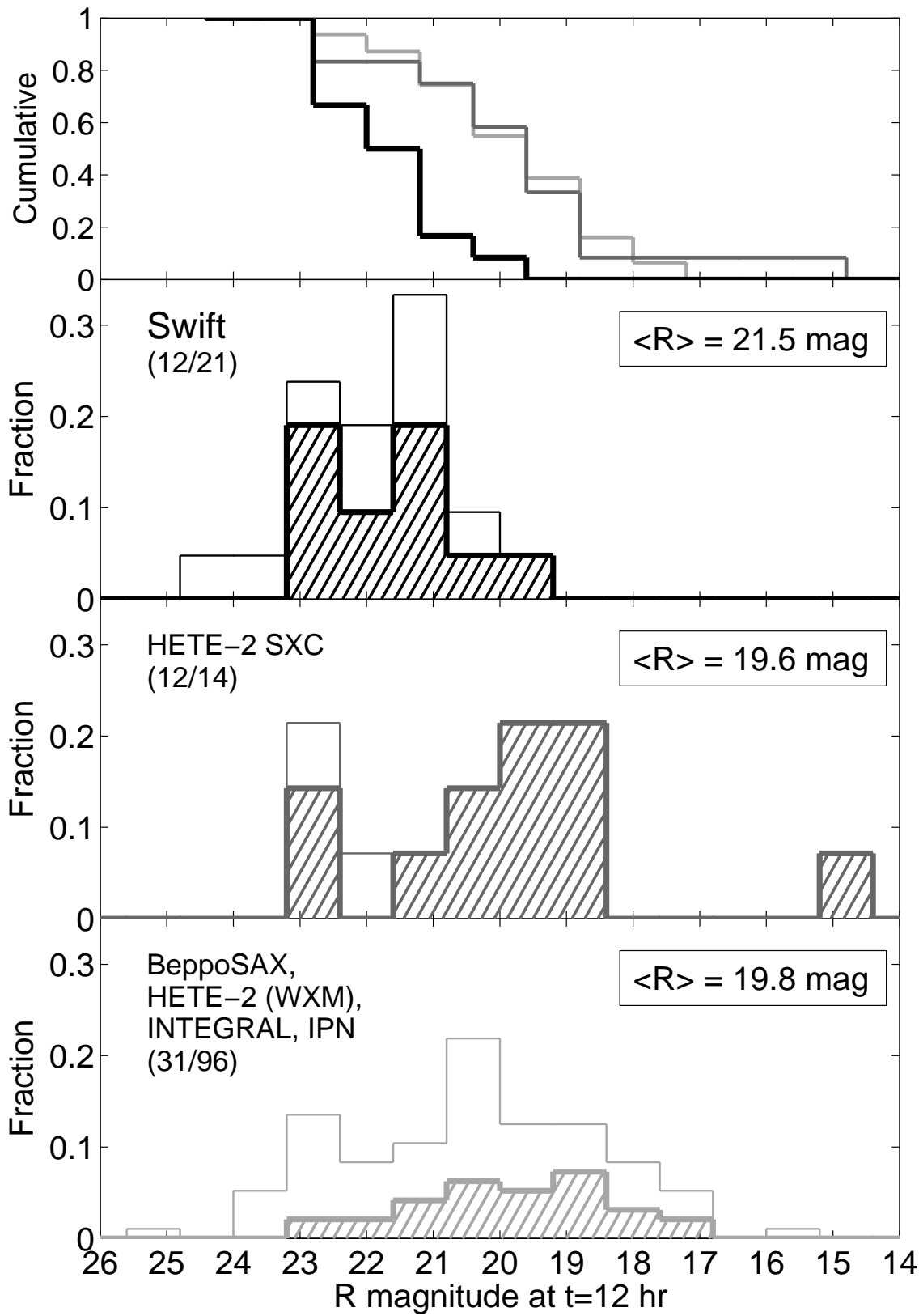,width=6.0in}}
\caption{Histograms of optical $R$-band magnitudes (corrected for
Galactic extinction) extrapolated to a common time of 12 hours after
the burst for the \swift, SXC, and BWI samples.  Shaded regions are
detections, while thin histograms include all upper limits.  The mean
$R$ magnitude of the detections in each sample is given as $\langle
R\rangle$.  The top panel shows the cumulative distributions.  The
afterglows of \swift\ bursts are fainter than those of bursts detected
in previous missions.  This is primarily the result of accurate and
rapid localizations, which have allowed us to increase the recovery
rate through the detection of fainter objects, and the lower threshold
of \swift (Figure~\ref{fig:lognlogs}), which results in detection of
fainter bursts.  The SXC sample, with a detection rate of $\sim 85\%$
in the optical, contains many bright afterglows, suggesting a bias in
favor of bright bursts.
\label{fig:optical}}
\end{figure}

In Figure~\ref{fig:optical} we plot the $R$-band magnitudes for the
three samples, normalized to $t=12$ hr.  We extrapolate (or
interpolate) to the fiducial time (and for near-IR afterglows to the
fiducial band) using the measured temporal and spectral slopes or by
conservatively assuming\footnotemark\footnotetext{The choice of
spectral and temporal indices is appropriate for the case of spherical
geometry, a constant density circumburst environment, an electron
power law index $p=2.2$, and a synchrotron cooling frequency
$\nu_c>\nu_R$.  This provides the most conservative decay rate: for a
Wind medium the dependence is $t^{-1.4}$, for $\nu_c<\nu_R$ it is
$t^{-1.15}$, and for the case of a collimated explosion it is
$t^{-p}\sim t^{-2.2}$.} $F_\nu\propto t^{-0.9}\nu^{-0.6}$.  As can be
seen from Figure~\ref{fig:optical}, the \swift\ afterglows, with a
mean $\langle R\rangle=21.5$ mag, are fainter relative to the SXC and
BWI samples by about $1.7$ magnitudes.  In fact, while $40\%$ of all
afterglows from past missions had $R<19$ mag at $t=12$ hr, not a
single \swift\ burst falls in that bright category.

\begin{figure}
\centerline{\psfig{file=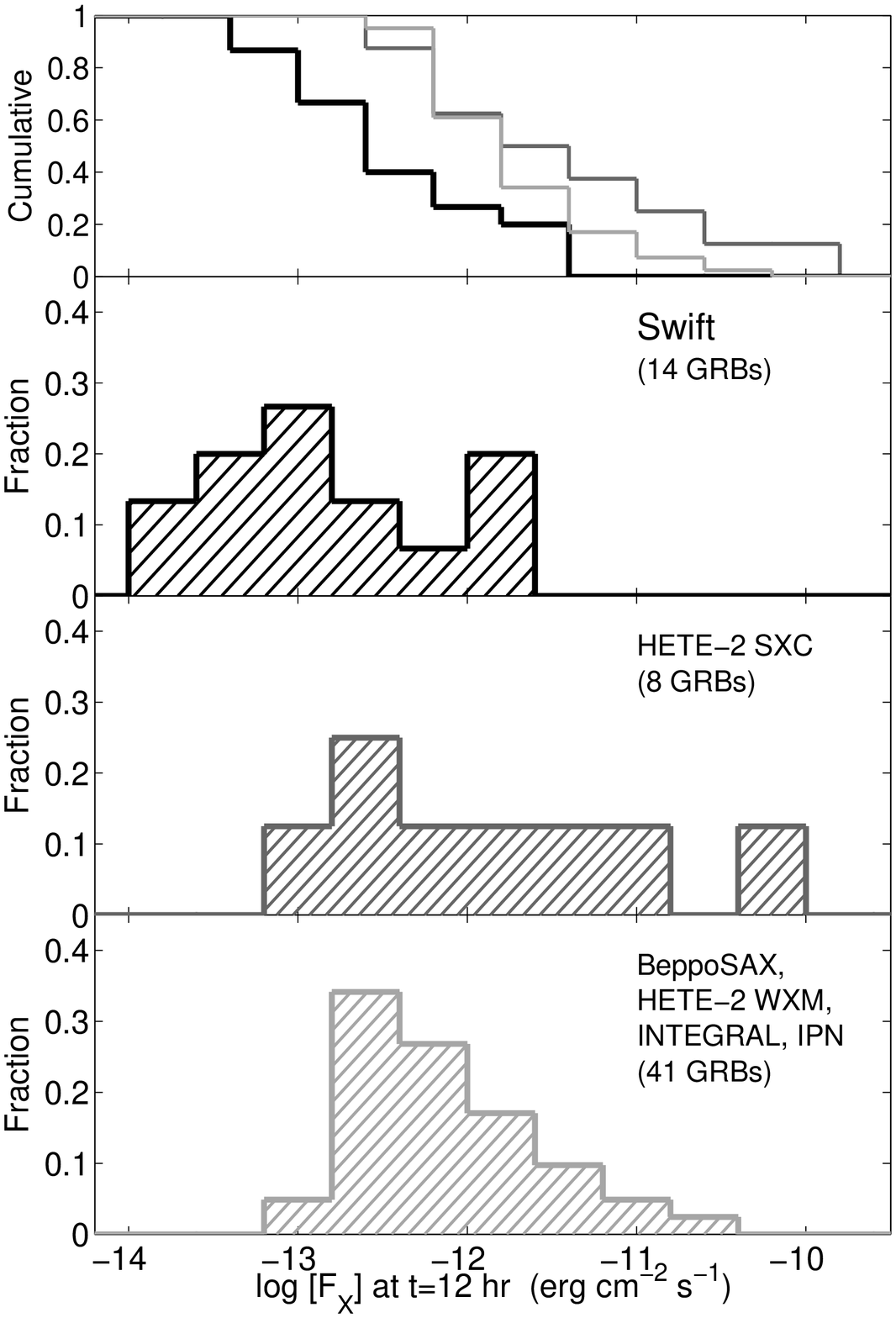,width=6.0in}}
\caption{Histograms of X-ray fluxes extrapolated to a common time of
12 hours after the burst for the \swift, SXC, and BWI samples.  The
top panel shows a cumulative distribution.  The distribution for
pre-\swift\ bursts is from \citet{bkf03} and \citet{bfk+05}.  The
X-ray afterglows of \swift\ bursts are fainter than those of bursts
detected in previous missions.  Since the past recovery rate was
already about $90\%$, this suggests that the lower trigger threshold
of \swift\ compared to BeppoSAX and HETE-2 is giving rise to fainter
and higher redshift bursts.
\label{fig:xray}}
\end{figure}

We carry out a similar exercise for the X-ray afterglow emission
(Figure~\ref{fig:xray}).  The X-ray fluxes at $t=12$ hr are estimated
using the measured decay indices (when available), or the mean value
based on all bursts, $\langle\alpha_X\rangle=-1.3$; for XRT fluxes
measured in the first hour we use a more conservative $\alpha_X=-1$.
As with the optical/near-IR afterglows, the X-ray afterglows of
\swift\ bursts are fainter relative to those of the other two samples 
by about a factor of five.

\begin{figure}
\centerline{\psfig{file=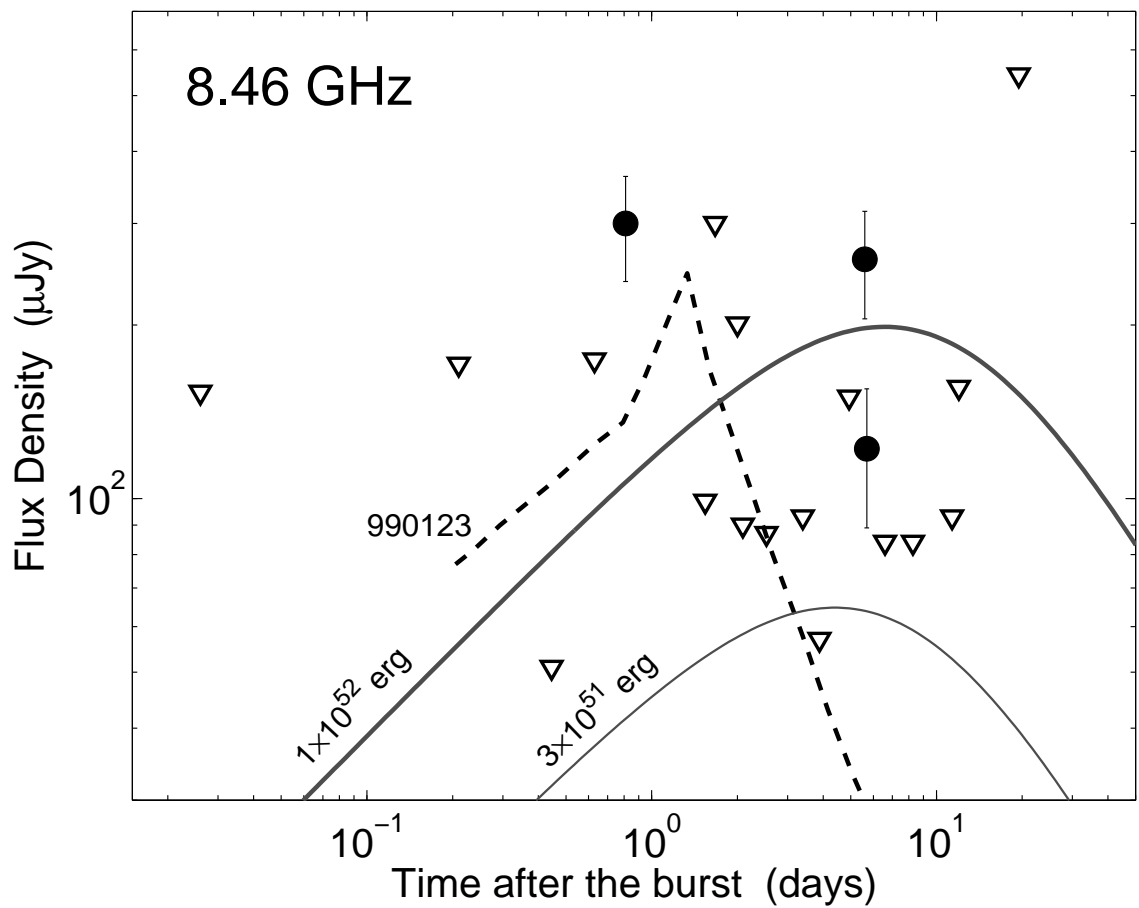,width=6.0in}}
\caption{Detections (circles) and upper limits (triangles) in the 
radio for \swift\ bursts.  Also shown are the radio light curve of
GRB\,990123 which was dominated by reverse shock emission, and
synthetic light curves of the forward shock emission from a burst with
typical parameters ($n_0=3$ cm$^{-3}$, $\epsilon_e=0.1$, and
$\epsilon_B=0.01$) placed at $z=2$ and with blastwave energies of
$1\times 10^{52}$ erg (thick line) and $3\times 10^{51}$ erg (thin
line).  Clearly, if \swift\ bursts typically have fainter afterglows
(e.g., less energy), as indicated by their optical and X-ray fluxes,
the majority may be too faint to detect at the VLA sensitivity.
\label{fig:radio}}
\end{figure}

Our comprehensive radio follow up of thirteen \swift\ bursts led to
the detection of only three (GRBs 050315, GRB 050401, GRB 050416a;
Figure~\ref{fig:radio}).  This is comparable to the $30\%$ recovery
fraction of the BWI sample, but is lower than the $55\%$ recovery of
the SXC sample.

\begin{figure}
\centerline{\psfig{file=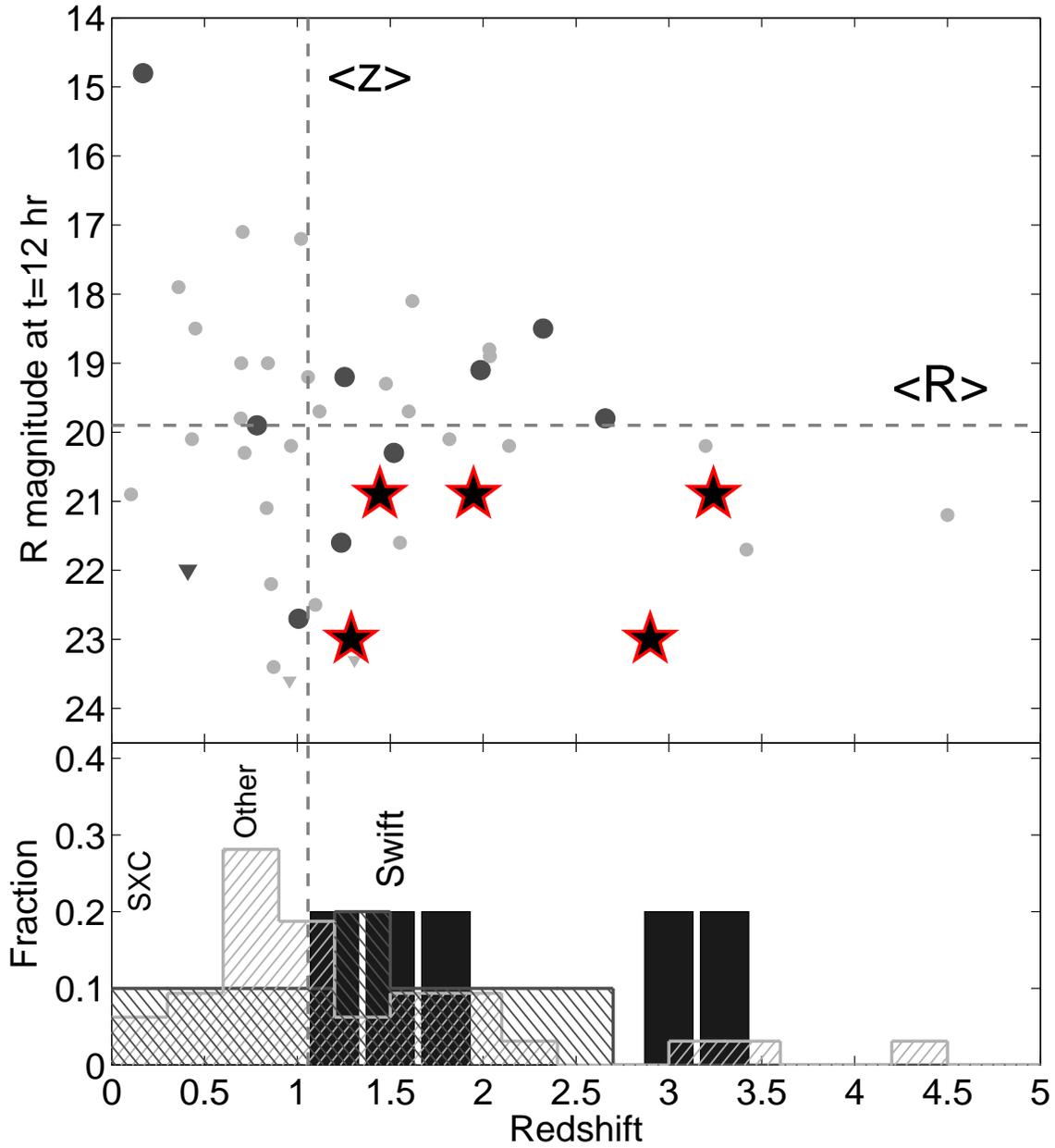,width=6.0in}}
\caption{Optical $R$-band magnitudes (corrected for Galactic
extinction) extrapolated to a common time of 12 hours after the burst,
plotted against redshift.  Light gray circles designate BWI bursts,
dark gray circles are SXC bursts, and stars are \swift\ bursts.  The
dashed lines designate the median magnitude and redshift for all
pre-\swift\ bursts.  The higher redshifts of \swift\ bursts are the
result of deeper and faster searches which uncover fainter afterglows,
as well as the lower threshold of \swift.
\label{fig:zopt}}
\end{figure}

The redshift distribution of \swift\ bursts differs from that of the
BWI sample, which peaks at $z\sim 0.8$ (Figure~\ref{fig:zopt}).  In
fact, the flat distribution of \swift\ bursts is similar to that of
the SXC sample.  It is interesting to note that the redshifts of the
(admittedly small) \swift\ sample are all beyond $1.25$, while the
median redshift for pre-\swift\ bursts is 1.05.

To summarize, \swift\ bursts, relative to both the BWI and SXC
samples, have fainter X-ray and optical afterglows.  \swift\ and the
BWI sample have a similar and low ($30\%$) recovery fraction for
radio afterglows.  We explain these findings as follows.  The BWI
sample with its moderate localization precision favored brighter
afterglows, which in turn favored lower redshift events.  The
\swift/XRT positions, combined with a greater trigger sensitivity
(Figure~\ref{fig:lognlogs}), allow us to increase the detection 
fraction and hence to find fainter
and higher redshift events.  The low rate of radio recovery is
primarily a result of the VLA sensitivity (which is lower relative to
the optical/near-IR bands).  The accurate and faster positions
available from \swift\ do not help to increase the radio detection
fraction.  In fact, the fainter afterglows of \swift\ bursts may in
the long run result in a lower recovery fraction in the radio
(Figure~\ref{fig:radio}).

The statistics of the SXC sample, however, are puzzling.  The high
afterglow detection fraction for SXC bursts, and their flat redshift
distribution (unlike the the BWI sample), can be attributed to better
localizations.  However, with a higher recovery rate we would expect
the SXC afterglows to be fainter than those of the BWI sample, while
they are in fact just as bright.  Similarly, the recovery fraction of
radio afterglows, which is set by the VLA threshold, is expected to be
comparable to that of the BWI and \swift\ samples; it is instead
significantly higher.  

There are two possible explanations.  Either the SXC sample is
suffering severely from Poisson statistics (though this is unlikely),
or the sample is biased to brighter afterglows.  We speculate that
soft X-ray contribution from the very early afterglow, or excess soft
X-ray emission in the prompt phase (e.g., \citealt{vsb+04}) may
enhance detection by the SXC, and possibly give rise to a sample with
brighter afterglows.  It appears that the SXC sample may not serve as
a proxy for the true distribution of afterglow properties, including
the fraction of dark bursts.

\section{Discussion and Ramifications}
\label{sec:disc}

\swift\ has been in orbit for six months and has localized 32 bursts
as of the end of April 2005.  In this paper we presented the results
of our ground-based optical/near-IR and radio follow up programs of
the 21 bursts with accurate positions from the XRT, including the
determination of three redshifts.

There is high expectation amongst astronomers that \swift, over the
remaining 30 months of its prime phase, will help address several
major questions: Are \swift\ bursts representative of the overall GRB
population?  How do GRBs evolve with redshift?  Is there a large
population of nearby events like GRBs 980425 and 031203?  Separately,
many astronomers are poised to exploit the brilliance of the
afterglows to investigate the star-formation history of the Universe,
probe the intergalactic medium in the early Universe, and investigate
the build up of elements in the disks of galaxies.  For these
astronomers the following questions are of prime importance: What is
the fraction of dark bursts, and how many of these lie at high
redshift ($z>6$)?  Will the afterglows be bright enough to undertake
high resolution optical/near-IR spectroscopy?  What is the annual
\swift\ rate of such interesting bursts?  We believe that our
investigation of the current sample has started to address some
aspects of the above questions.

\begin{figure}
\centerline{\psfig{file=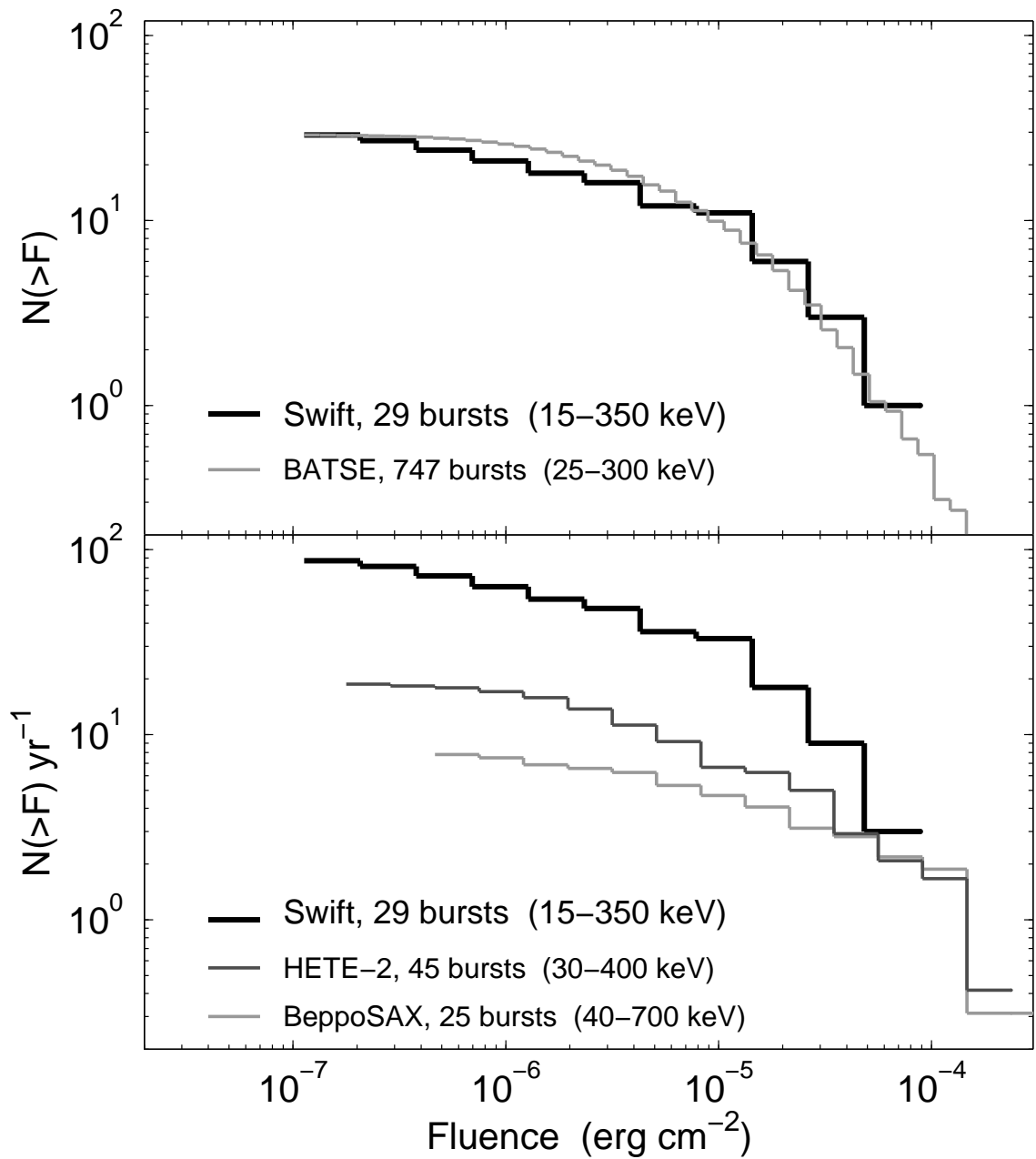,width=6.0in}}
\caption{{\it Top:} ${\rm log}N/{\rm log}F$ for \swift\ bursts with 
published fluences compared to a sample of BATSE bursts for which a
fluence measurement is available in all four channels and the error is
smaller than half of the measured value.  The BATSE fluence
distribution is normalized by the relative number of bursts.  The
similarity of the distributions and thresholds explain the lack of
local, low luminosity bursts in the current \swift\ sample, as well as
the lower event rate compared to pre-launch estimates (90 vs.~150 per
year, respectively).  {\it Bottom:} The same \swift\ sample but
compared to BeppoSAX and HETE-2 bursts \citep{aft+02,slg+04,ggf+05}.
For HETE-2 we included only GRBs and X-ray rich GRBs.  Each sample is
normalized by the approximate detection rates per year.  Clearly,
\swift\ has a lower threshold and this results primarily in an 
increase in the number of faint bursts.
\label{fig:lognlogs}} 
\end{figure}

To start with, the observed \swift\ rate is about $80-90$ bursts per
year.  This is less than the 150 bursts per year estimated from
pre-launch simulations, which assumed a threshold of five times better
than that of BATSE \citep{fmp+04}.  The smaller rate is consistent
with the fact that the fluences and peak count rates of the observed
\swift\ events are in fact similar to those that triggered BATSE, as
are the resulting threshold and the ${\rm log}N/{\rm log}S$
distribution (Figure~\ref{fig:lognlogs}).

Next, there is at the present no indication of a large sample of
nearby ($z\lesssim 0.2$) low luminosity events.  As noted earlier, the
lowest measured redshift is $z\approx 1.29$. More importantly, there
is no evidence of bright galaxies\footnotemark\footnotetext{In the
host galaxy sample for pre-\swift\ GRBs, all objects at $z<0.5$ have
$R\lesssim 22$ mag, and a mean of $20.6$ mag.  Similar hosts should
have been detected in most of the afterglow searches of \swift\ XRT
positions (Table~\ref{tab:obs}).} in the XRT error circles.  The lack
of low redshift objects is not surprising given that \swift's
threshold is similar to that of BATSE.  In the BATSE sample, even when
including the faint, non-triggered bursts, the limit on a contribution
from a spatially homogeneous, local population is about $7\%$ ($90\%$
confidence; \citealt{klk+00}).

In addition, the \swift\ afterglows are fainter and their redshifts
are higher.  The rapid response of a variety of telescopes to \swift\
events, and their accurate positions, means that the \swift\ sample
has the least bias (relative to the previous samples).  We therefore
conclude that (i) the true GRB population peaks at a higher redshift
than indicated by pre-\swift\ bursts, $z\gtrsim 2$, and (ii) the
typical optical and X-ray afterglows are faint: at $t=12\,$hr, $R\sim
21.5$ mag and $F_X\sim 3\times 10^{-13}$ erg cm$^{-2}$ s$^{-1}$,
respectively.

A number of ramifications follow from the above discussion.  First,
the fainter afterglows mean that it is critical that moderate (and
large) telescopes be pressed into service to identify optical/near-IR
afterglows; the expected signal level (at $t=15$ min) of $R\sim 17.7$
mag, $J\sim 16.6$ mag, and $K\sim 15.2$ mag is difficult to achieve on
small robotic telescopes.  In addition, the combination of higher
redshifts and fainter afterglows strongly favor longer wavelength
observations ($RIz$ bands versus $UBV$); indeed, a simple way to
improve the current low detection fraction by the UVOT (5/16) is to
observe in only the $V$ band (unless the afterglow is noted to be
bright).  Finally, near-IR observations are essential to distinguish
dusty events from high redshift events.

So far we have tacitly assumed that \swift\ events are representative of
the true sample, and the afterglow and redshift distributions are simply
due to a lower threshold and better localizations than BeppoSAX and
HETE-2.  However, a possible bias arises from the the softer band of the
BAT ($15-150$ keV) compared to the BeppoSAX GRBM ($40-700$ keV) and
HETE-2 FREGATE ($6-1000$ keV) instruments.  \citet{aft+02} and
\citet{slg+04} show that there is a correlation between the peak energy
of the prompt emission spectrum, $E_{\rm peak}$, and the fluence (or
peak flux).  This is an approximate relation but it appears to be obeyed
in the mean.  This means that the BAT is well suited for finding ``X-ray
rich GRBs'', which will therefore result in selection of bursts with a
lower fluence and peak flux.  This is potentially a significant bias in
the \swift\ sample.  Conversely, the softer triggering band of the BAT
means that the detection of short-hard bursts will be diminished.  This
may explain why \swift\ has localized only a single short-hard burst out
of the sample of 32 bursts.

The aggregate effect of \swift's trigger threshold, energy band, and
localization accuracy has resulted in a sample that is dominated by
faint optical/near-IR and X-ray afterglows with seemingly higher
redshifts than indicated by past missions.  The current recovery rate
suggests that as many as $1/3$ of all \swift\ bursts may be optically
dark.  Therefore, while a local population of low energy bursts does
not contribute significantly to the sample, the possibility that
\swift\ will uncover a larger population of high redshift bursts than
past missions remains open.  Follow-up efforts and resources should be
focused on this possibility.

\acknowledgements 
We thank Neil Gehrels, Don Lamb, and Tsvi Piran for useful discussions. 
E.B. is supported by NASA through Hubble Fellowship grant HST-01171.01 
awarded
by the Space Telescope Science Institute, which is operated by AURA,
Inc., for NASA under contract NAS 5-26555.  A.G. is supported by NASA
through Hubble Fellowship grant HST-01171.01 awarded by the Space
Telescope Science Institute, which is operated by AURA, Inc., for NASA
under contract NAS 5-26555.  B.P.S. and B.A.P. are supported by ARC
grant DP0559024.  Additional support was provided by NSF and NASA
grants.


\clearpage
\begin{deluxetable}{llllcrcllllcl}
\tablecolumns{13}
\tabcolsep0.05in\footnotesize
\tablewidth{0pc}
\tablecaption{Gamma-Ray Burst and Afterglow Properties
\label{tab:obs}}
\tablehead {
\colhead {GRB}       		&
\colhead {$z$}		        &
\colhead {$\delta t_{\rm opt}$} &
\colhead {Telescope}  		&
\colhead {Filter}     		&
\colhead {Mag.}  		&
\colhead {$\delta t_{\rm rad}$} &
\colhead {$F_{\rm \nu,rad}$}    &
\colhead {$\delta t_X$} 	&
\colhead {$F_X$}    		&
\colhead {$F_\gamma$}           &
\colhead {$P_\gamma$}           &
\colhead {Refs.}                \\
\colhead {} 	  		    &
\colhead {}	 	            &
\colhead {(hr)}       		    &      
\colhead {}		       	    &      
\colhead {}		 	    &
\colhead {}   			    &
\colhead {(d)} 			    &
\colhead {($\mu$Jy)}       	    &       
\colhead {(s)}       		    &      
\colhead {(erg cm$^{-2}$ s$^{-1}$)} &
\colhead {(erg cm$^{-2}$)}          &
\colhead {(cnts cm$^{-2} s^{-1}$)}  &
\colhead {}
}
\startdata
041223  & \nod  & 14.4 & LCO40      & r & 20.81   & \nod & \nod        & $1.63\times 10^4$ & $6.5\times 10^{-12}$ & $5.0\times 10^{-5}$ & 7.5  & 1,2898 \\
050117a & \nod  & 14.6 & P200/WIRC  & K & $>18.8$ & 1.54 & $<99$       & $193$             & $1.8\times 10^{-8}$  & $1.7\times 10^{-5}$ & 0.9  & 1,2962 \\
050124  & \nod  & 24.5 & Keck/NIRC  & K & 19.66   & 4.93 & $<150$      & $2.54\times 10^4$ & $2.2\times 10^{-12}$ & $2.1\times 10^{-6}$ & 6.8  & 1,2973 \\
050126  & 1.290 & 4.32 & Keck/NIRC  & K & 19.45   & 2.09 & $<90$       & $200$             & $2.5\times 10^{-11}$ & $2.0\times 10^{-6}$ & 0.4  & 1,2987 \\
050128  & \nod  & 11.4 & Faulkes    & R & $>20.5$ & 11.3 & $<93$       & $873$             & $2.6\times 10^{-12}$ & $4.5\times 10^{-6}$ & 4.6  & 2991,2992,3001,3011 \\
050215b & \nod  & 9.00 & P60        & R & $>20.5$ & 3.39 & $<93$       & $5.7\times 10^3$  & few$\times 10^{-13}$ & $4.5\times 10^{-7}$ & \nod & 3032,3034,3035,3053,3066 \\
        &       & 9.76 & UKIRT/UFTI & K & 20.23   &      &             &                   &                      &                     &      & 3028,3031 \\
050219a & \nod  & 2.05 & MJUO 0.6-m & R & $>20.5$ & \nod & \nod        & \nod              & \nod                 & $9.4\times 10^{-6}$ & 5.5  & 3038,3041 \\
        & \nod  & 17.7 & LCO40      & I & $>21.5$ &      &             &                   &                      &                     &      & 3048 \\
050219b & \nod  & 4.32 & VLT/FORS2  & R & $>23.0$ & \nod & \nod        & \nod              & \nod                 & $2.3\times 10^{-5}$ & 26   & 3044,3064 \\
        & \nod  & 5.21 & LCO100     & K & 19.5    &      &             &                   &                      &                     &      & This paper    \\
050223  & \nod  & 4.10 & PROMPT     & R & $>21.2$ & \nod & \nod        & $7.67\times 10^4$ & $4.5\times 10^{-14}$ & $7.4\times 10^{-7}$ & 0.8  & 3055,3067,3109 \\
        &       & 5.22 & LCO40      & R & $>21.6$ &      &             &                   &                      &                     &      & This paper    \\
050306  & \nod  & 50.5 & TNG        & R & 23.0    & 8.27 & $<84$       & \nod              & \nod                 & $1.9\times 10^{-5}$ & 2.4  & 3089,3092      \\
050315  & 1.950 & 11.6 & Mag./LDSS3 & R & 20.9    & 0.81 & $300\pm 62$ & \nod              & \nod                 & $4.2\times 10^{-6}$ & 2.5  & 3098,3100,3101,3102,3105 \\ 
050318  & 1.444 & 8.15 & Mag./IMACS & R & 20.6    & \nod & \nod        & $7.7\times 10^3$  & $7.0\times 10^{-12}$ & $2.1\times 10^{-6}$ & 3.8  & 3112,3114,3122,3134 \\
050319  & 3.240 & 8.70 & RTT150     & R & 20.14   & 0.63 & $<174$      & \nod              & \nod                 & $8.0\times 10^{-7}$ & 1.7  & 3119,3127,3132      \\
050326  & \nod  & 6.90 & MJO 0.6-m  & I & $>20.3$ & \nod & \nod        & $5.4\times 10^4$  & $7.4\times 10^{-13}$ & $1.9\times 10^{-5}$ & 17   & 3145,3151,3293 \\
050401  & 2.900 & 0.96 & SSO40      & R & 20.3    & 5.69 & $122\pm 33$ & $1.9\times 10^4$  & $3.8\times 10^{-12}$ & $1.9\times 10^{-5}$ & 14   & 3163,3164,3179,3187 \\
050406  & \nod  & 7.80 & Mag./LDSS3 & R & 22.0    & \nod & \nod        & $3.8\times 10^4$  & $6.7\times 10^{-14}$ & $9.0\times 10^{-8}$ & 3.2  & 3183,3184,3185 \\ 
050410  & \nod  & 4.45 & ARIES      & R & $>20.5$ & 0.21 & $<171$      & \nod              & \nod                 & $6.9\times 10^{-6}$ & 2.0  & 3219,3223,3226 \\
        & \nod  & 15.6 & P60        & i & $>21.5$ &      &             &                   &                      &                     &      & 3231    \\
050412  & \nod  & 0.22 & P60        & R & $>20.0$ & 3.88 & $<57$       & $5.0\times 10^3$  & $3.9\times 10^{-12}$ & $2.1\times 10^{-6}$ & 0.8  & 3242,3251,3253,3277 \\
        &       & 0.83 & LCO100     & R & $>22.4$ &      &             &                   &                      &                     &      & This paper \\
050416a & \nod  & 3.50 & SSO 2.3-m  & R & 21.7    & 5.58 & $260\pm 55$ & $4.3\times 10^4$  & $2.8\times 10^{-13}$ & $3.8\times 10^{-7}$ & 4.8  & 3266,3273,3275,3318 \\
050416b & \nod  & 1.10 & Mag./IMACS & R & $>24.0$ & \nod & \nod        & \nod              & \nod                 & $2.1\times 10^{-6}$ & 7.9  & 3282,3284      \\
050421  & \nod  & 4.62 & P60        & R & $>22.0$ & 0.44 & $<102$      & $837$             & $6.4\times 10^{-13}$ & $1.8\times 10^{-7}$ & 0.5  & 3299,3301,3305,3308 \\
        &       & 1.70 & TNG        & K & $>18.6$ &      &             &                   &                      &                     &      & 3300 \\ 
        &       & 9.30 & MAGNUM     & J & $>20.3$ &      &             &                   &                      &                     &      & 3313 \\ \hline
050408$^{a}$ & 1.236 & 3.70 & RTT150  & R & 20.94 & 2.53 & $<87$       & $2.1\times 10^4$  & $6.2\times 10^{-12}$ & $3.3\times 10^{-6}$ & \nod & 3189,3191,3201,3234,3262
\enddata
\tablecomments{Prompt emission and afterglow properties for all \swift\ 
bursts with XRT positions and ground-based follow-up as of the end of
April 2005.  The columns are (left to right): (i) GRB name, (ii)
redshift, (iii) time of optical/near-IR observation, (iv) telescope,
(v) filter, (vi) optical/near-IR magnitude, (vii) time of radio
observation, (viii) radio flux at 8.46 GHz, (ix) time of X-ray
observation, (x) X-ray flux, (xi) $\gamma$-ray fluence, (xii)
$\gamma$-ray peak flux, and (xiii) references (1.~\citet{bfk+05}, all
other are GCN numbers:
\citet{gcn2898}, \citet{gcn2962}, \citet{gcn2973}, \citet{gcn2987}, 
\citet{gcn2991}, \citet{gcn2992}, \citet{gcn3001}, \citet{gcn3011}, 
\citet{gcn3032}, \citet{gcn3034}, \citet{gcn3035},
\citet{gcn3053}, \citet{gcn3066}, \citet{gcn3028}, \citet{gcn3031}, 
\citet{gcn3038}, \citet{gcn3041}, \citet{gcn3048}, \citet{gcn3044}, 
\citet{gcn3064}, \citet{gcn3055}, \citet{gcn3067}, \citet{gcn3109}, 
\citet{gcn3089}, \citet{gcn3092}, \citet{gcn3098}, \citet{gcn3100}, 
\citet{gcn3101}, \citet{gcn3102}, \citet{gcn3105}, \citet{gcn3112}, 
\citet{gcn3114}, \citet{gcn3122}, \citet{gcn3134}, \citet{gcn3119},
\citet{gcn3127}, \citet{gcn3132}, \citet{gcn3145}, \citet{gcn3151}, 
\citet{gcn3293}, \citet{gcn3163}, \citet{gcn3164}, 
\citet{gcn3179}, \citet{gcn3187}, \citet{gcn3183}, \citet{gcn3184}, 
\citet{gcn3185}, \citet{gcn3219}, \citet{gcn3223}, \citet{gcn3226}, 
\citet{gcn3231}, \citet{gcn3242}, \citet{gcn3251}, \citet{gcn3253}, 
\citet{gcn3266}, \citet{gcn3273}, \citet{gcn3275}, \citet{gcn3318}, 
\citet{gcn3282}, \citet{gcn3284}, \citet{gcn3299}, \citet{gcn3301}, 
\citet{gcn3305}, \citet{gcn3308}, \citet{gcn3300}, \citet{gcn3313}, 
\citet{gcn3189}, \citet{gcn3191}, \citet{gcn3201}, \citet{gcn3234}, 
\citet{gcn3262}).  $^a$ HETE-2 SXC burst.}
\end{deluxetable}

\clearpage

\clearpage

\clearpage

\clearpage

\clearpage

\clearpage

\clearpage

\clearpage

\end{document}